\documentclass{emulateapj} \usepackage{apjfonts} \usepackage{mathletters} 
\usepackage{amstext} 
\usepackage{amsmath}

\begin{document}

\submitted 
{Submitted to ApJ}
\journalinfo 
{Submitted to ApJ}

\shorttitle{Magnetic Driving of Relativistic Outflows in AGNs. I.}
\shortauthors{{VLAHAKIS} AND {K\"ONIGL}}

\title{Magnetic Driving of Relativistic Outflows in Active Galactic Nuclei.
\\
I. Interpretation of Parsec-Scale Accelerations}

\author{Nektarios Vlahakis}
\affil{Section of Astrophysics, Astronomy \& Mechanics,
Department of Physics, University of Athens, 15784 Zografos Athens, Greece
\\ vlahakis@phys.uoa.gr}
\and
\author{Arieh K\"onigl}
\affil{Department of Astronomy \& Astrophysics and Enrico Fermi
Institute, University of Chicago, 5640 S. Ellis Ave., Chicago, IL 60637
\\ arieh@jets.uchicago.edu}

\begin{abstract}
There is growing evidence that relativistic jets in active
galactic nuclei undergo extended (parsec-scale) acceleration. We
argue that, contrary to some suggestions in the literature, this
acceleration cannot be purely hydrodynamic. Using exact semianalytic
solutions of the relativistic MHD equations, we demonstrate that
the parsec-scale acceleration to relativistic speeds inferred in sources
like the radio galaxy NGC 6251 and the quasar 3C 345 can be
attributed to magnetic driving. Additional observational
implications of this model will be explored in future papers in
this series.
\end{abstract}


\section{Introduction}\label{introduction}
Magnetic acceleration and collimation has long been thought to
be the underlying mechanism responsible for the similar
manifestations of cosmic jets in such diverse systems as young
stellar objects and active galactic nuclei (AGNs)
\citep[e.g.,][]{K86,P93,S96,L00}. 
Although some AGN jets have long been known to exhibit
apparent superluminal motions, with inferred (terminal) bulk Lorentz
factors in the blazar class of sources of $\gamma_\infty
\lesssim 10$ (but exceeding 40 in some cases; e.g.,
\citealt{J01}), most models to date have concentrated on the
nonrelativistic regime. However, two recent discoveries --- the
detection of apparent superluminal
motions in certain Galactic black-hole binaries (the so-called
microquasars; e.g., \citealt{MR99}),
from which mildly relativistic bulk velocities have been
deduced, and the inferred association of gamma-ray bursts (GRBs)
with ultrarelativistic ($\gamma_\infty \gtrsim 10^2$), highly
collimated outflows \citep[e.g.,][]{P99} --- have highlighted the 
strong similarities among the various types of relativistic jet
sources \citep[e.g.,][]{GC02} and have refocused attention on
the question of their origin. Although the interpretation of
relativistic outflows and the generalization of magnetohydrodynamics (MHD) to the
relativistic regime present several distinct challenges, the
prevailing view has been that magnetic driving is the common
underlying mechanism also in this case \citep[e.g.,][]{B02}.

However, this interpretation of relativistic jets is by
no means universal. One example is provided by the radio galaxy
NGC 6251, in which \citet{S00} inferred a bulk acceleration from
$V\approx 0.13\, c$ to $V\approx 0.42\, c$ on sub-parsec scales.
This behavior was attributed by \citet*{MLF02} to a thermal acceleration of a
proton-electron plasma that is heated to a temperature $T
\approx 10^{12}\, {\rm K}$ in a region of radius $r \lesssim 0.03\, {\rm pc}$. We note, however,
that a thermally driven, purely hydrodynamic flow
typically undergoes the bulk of its acceleration over a distance
that is of the order of the size of the mass distribution that
initially confines it by its gravity, which 
in this case is much smaller than the radius of the apparent
acceleration zone (see \S~\ref{thermalaccel}).
Centrifugal driving \citep[e.g.,][]{BP82} --- the commonly invoked hydromagnetic
acceleration mechanism for nonrelativistic jets --- typically also
acts fairly rapidly and thus would similarly fail to account for the
large-scale acceleration inferred in NGC 6251. A possible
resolution of this puzzle is provided by the finding of \citet*{LCB92},
who were the first to generalize the ``cold'' radially self-similar MHD flows
of \citet{BP82} to the relativistic regime (see also
\citealt{C94}), that their solutions contain an extended magnetic
pressure-gradient acceleration region beyond the
classical fast-magnetosonic point
(a singular point of the Bernoulli equation), a behavior
that they ascribed to the action of a ``magnetic nozzle.'' It was subsequently shown by
\citet{V00} that a similar mechanism operates also in
nonrelativistic flows, but the effect is probably easier to discern
observationally in relativistic jets.

Vlahakis \& K\"onigl (2003a, hereafter VK) carried out a further generalization
by deriving ``hot,'' radially self-similar, relativistic MHD
solutions for trans-Alfv\'enic flows.\footnote{The
trans-Alfv\'enic solutions correspond to a dominant poloidal
magnetic field at the base of the flow. \citet{VK03b} derived
analogous solutions for super-Alfv\'enic jets, for which the
magnetic field at the base of the flow is predominantly
azimuthal. The latter configuration may be expected to apply
to inherently nonsteady outflows \citep[e.g.,][]{C95}. In this paper we adopt the
poloidal field configuration as the most appropriate modeling
framework for AGN jets, as has also been done by other workers \citep*[e.g.,][]{LPK03}.}
They showed that the magnetic field always guides and collimates
the flow, but that, if 
the specific enthalpy $\xi c^2$ is initially (subscript $i$) $\gg c^2$, 
then an extended {\em thermal} acceleration region can develop,
within which the flow is accelerated from $\gamma_i \approx 1$ to
$\gamma \approx \xi_i$. If the 
total energy-to-mass flux ratio $\mu c^2$ is $\gg \xi_i c^2$, 
corresponding to a Poynting
flux-dominated outflow, then the bulk of the acceleration is
magnetic and takes place downstream from this point. VK
demonstrated that the flow continues to be accelerated all the
way up to the {\em modified} fast-magnetosonic surface, which
is the locus of the fast-magnetosonic
singular points of the {\em combined} Bernoulli and transfield
equations and represents the true ``causality surface'' (or ``event horizon'') for the
propagation of fast waves. VK showed that this singular surface can
lie well beyond the classical fast-magnetosonic surface and
argued that this is the essence of the ``magnetic nozzle'' effect.

This is the first in a series of papers in which we apply the VK
formalism to the interpretation of relativistic jets in
AGNs.\footnote{We have previously concentrated on applications to
GRBs; see VK, where analogies among GRBs, AGNs,
and microquasars are discussed, as well as Vlahakis \&
K\"onigl (2001, 2003b) and \citet*{VPK03}.\label{grbref}}
Our aim is to model a variety of observational findings and
attempt to construct basic diagnostic tools for the study of such
jets. In this paper we focus on the extended-acceleration
signature of magnetically driven jets and use our solutions to model
the parsec-scale accelerations already indicated in a number of
relativistic jet sources. 
We first present arguments for why the  extended acceleration is
unlikely to have a thermal origin (\S~\ref{thermalaccel}). We then
consider magnetic jet models: after a brief review of the solution
methodology (\S~\ref{model}), we demonstrate (\S~\ref{magnetic})
that magnetic driving can account
for Sudou et al.'s observations of NGC 6251 as well for the parsec-scale
acceleration to $\gamma_\infty \gtrsim 10$ inferred in
superluminal blazar jets like 3C 345 \citep{U97}. Our
conclusions are given in \S~\ref{conclusion}.

\section{Can the Jets be Driven Thermally?}\label{thermalaccel}
The sub--parsec-scale kinematics of the NGC 6251 jet was deduced by
\citet{S00} after they discovered the counterjet in this powerful radio galaxy.
By applying a relativistic beaming model to the measured
variation in the jet--counterjet intensity ratio,
they inferred that the outflow is accelerated from $\sim 0.13\, c$ 
at $r\approx 0.53\ {\rm pc}$ to
$\sim 0.42\, c$ at $r\approx 1.0\ {\rm pc}$.\footnote{Using a
more accurate determination of
the distance to NGC 6251, \citet{MLF02} took the radial range of
the acceleration region to be $0.30-0.57\ {\rm pc}$. 
Since the differences from the values adopted by
\citet{S00} have little impact on our arguments, we continue to
use the latter in our discussion.} 
We now argue that, contrary to the suggestion made in
\citet{MLF02}, the cause of this acceleration cannot be thermal
pressure driving.

Since the inferred speeds are not too close to $c$, 
it is sufficient to use nonrelativistic hydrodynamics.
In the absence of magnetic forces and assuming spherical
symmetry for simplicity,
the flow can be described as a Parker (\citeyear{P58}) wind, in which the
radial velocity $V$ is given as a function of the radius $r$ by
\begin{equation}\label{parker}
\frac{r}{V} \frac{dV}{dr}=
\frac{2 C_{\rm s}^2 - GM/r } {V^2- C_{\rm s}^2}  \,.
\end{equation}
Here $M$ is the mass of the central black hole,
$G$ is the gravitational constant, and 
$C_{\rm s}=\left(2\Gamma k_{\rm B} T / m_p\right)^{1/2}$ is
the sound speed of a  fully ionized hydrogen gas
(with $\Gamma$, $m_p$ and $k_{\rm B}$ being the adiabatic index, proton mass, and
Boltzmann's constant, respectively).
In the best-fit model of \citet{MLF02} (which is consistent 
with the limits set by the radio observations of
\citealt{J86}), the base temperature (at $r_i=0.026\ {\rm pc}$)
is $T_i=10^{12}\ {\rm K}$. For these values of $r$ and $T$, and
with $M\approx 6\times 10^{8}\ M_\odot$ \citep{FF99},
we find that $2 C_{\rm s}^2$ is $\gg GM/r$ and hence that the
flow at $r_i$ would already be supersonic with
$V > C_{\rm s} \approx 0.55\, c$. By the time such a flow reached
the scales mapped by \citet{S00}, its velocity would be
significantly larger  --- and its acceleration substantially smaller
--- than what has been inferred from the observations.
 
Extended acceleration has also been indicated in superluminal
blazar jets. For example, VLBI images of the quasar 3C 345 have shown that
the jet component speeds increase with separation from the core
\citep[e.g.,][]{Z95,LZ99}. In particular, in the case of the C7 component, \citet{U97}
combined a VLBI proper-motion measurement with an inference of the
Doppler factor from a synchrotron self-Compton calculation
to deduce an acceleration from $\gamma \sim 5$ to $\gamma \gtrsim 10$
over a (deprojected) distance range (measured from the core) of
$\sim 3-20\ {\rm pc}$.

Although blazar jets are also sometimes modeled in terms of a purely hydrodynamic
acceleration \citep[e.g.][]{GM98},
one can show quite generally that in this case, too, thermal driving alone cannot
account for the observations. A variety of
arguments (summarized, e.g., in \citealt{GC02}) indicate that
protons are the dynamically dominant component in many AGN
jets. Energy conservation in such outflows implies
$\gamma_\infty / \gamma_i\approx 1+2 [\Gamma/ (\Gamma-1)] (k_{\rm
B} T_i / m_p c^2)$. Even if the initial temperature were as high
as $\sim 10^{12}\ {\rm K}$, the terminal
Lorentz factor would still be $\lesssim 2$ --- much smaller
than the values typically inferred from the observed superluminal motions.
An additional argument can be made on the basis of the observed
acceleration rate, which in a purely hydrodynamic model is
determined by the radial dependence of the external pressure
that provides lateral confinement of the
jet. In a simplified picture of a one-dimensional
pressure distribution that scales with distance $z$ from the
origin as $z^{-\alpha}$, the Lorentz factor of an adiabatic, supersonic jet
is predicted to increase as $z^{\alpha/4}$ \citep[e.g.,][]{BR74}. 
Applying this picture to the 3C 345 data reported in \citet{U97}, 
one infers $\alpha \sim 1.5$ on parsec scales. It is, however, unclear how 
a pressure distribution of this type could arise in a natural way in an AGN.

The most likely alternative in both of these cases is magnetic acceleration, which we
consider in the remainder of this paper.

\section{Semianalytic Solutions of the ``Hot''
Relativistic MHD Equations}\label{model}

VK considered the full set of special-relativistic MHD
equations, allowing both the bulk and the random speeds to be
relativistic. They assumed ideal MHD, axisymmetry, no explicit time
dependence, and a polytropic equation of state (with the pressure $P$
scaling with the rest-mass density $\rho_0$ as $P\propto
\rho_0^\Gamma$, where the adiabatic index is taken to
be 5/3 or 4/3 depending, respectively, on whether the pressure is dominated by ``cold''
protons and electrons or by ``hot'' electron-positron pairs and
radiation). Under these assumptions, the MHD equations can be
partially integrated to yield several field-line constants
(with the field
line being identified by the poloidal magnetic
flux function $A$): the total specific angular momentum $L(A)$,
the field angular velocity $\Omega(A)$, the magnetization
parameter $\sigma_{\rm M}(A)$ (with the mass-to-magnetic flux ratio
given by ${A\Omega^2}/{\sigma_{\rm M} c^3}$), the adiabat $Q(A)=
P/ \rho_0^{\Gamma}$, and the 
total energy-to-mass flux ratio 
$\mu(A) c^2 =\xi \gamma c^2 - (c/4 \pi)(E B_\phi/\gamma \rho_0 V_p)$ 
(where $E$ is the electric field amplitude, $B_\phi$ is the azimuthal field
component, and $V_p$ is the poloidal velocity component).
For the adopted equation of state, the specific enthalpy is
given by $\xi c^2 = c^2+ [\Gamma/(\Gamma-1)] (P/\rho_0)$.

The transfield force-balance equation is integrated under the
most general {\em ansatz} for radial self-similarity [in
spherical coordinates $(r\,,\theta\,,\phi)$)], in which the
shape $r(A\,, \theta)$ of a poloidal field line is given as a
product of a function of $A$ times a function of $\theta$:
$r={\cal F}_1(A) {\cal F}_2 (\theta)$ \citep[see][]{VT98}. Such
a separation of variables can be effected if ${\cal F}_1(A)
\propto A^{1/F}$, $L(A) \propto A^{1/F}$, $\Omega(A) \propto A^{-1/F}$,
$Q(A) \propto A^{-(F-2)(\Gamma-1)}$, and if $\mu(A)= const$ and
$\sigma_{\rm M}(A)  = const$. 
To obtain a solution, it is
necessary to specify seven boundary conditions as well as the
values of $\Gamma$ and $F$. The parameter $F$
controls the distribution of the poloidal current $I$: $2I/c = \varpi B_\phi
= A^{1-1/F} {\cal F}(\theta)$ [using cylindrical coordinates
($\varpi\,,\phi\,,z)$].
Close to the origin the field is
force-free, with ${\cal F}(\theta) \approx const$, which implies
$\varpi B_\phi \propto A^{1-1/F}$. For $F>1$, the current $|I| $
is an increasing function of $A$, corresponding to the current-carrying
regime (for which the poloidal current density is antiparallel to the
field). A solution with $F>1$ should provide a good
representation of the conditions near the axis of a highly
collimated flow. Conversely, a solution with $F<1$ corresponds to
the return-current regime (for which the poloidal current density is parallel to the
field) and may be most suitable at larger cylindrical
distances. VK showed that, for 
$F\gtrsim 1$, the modified fast surface is at infinity
and an initially Poynting-dominated flow ($\mu \gg \xi_i$) attains a rough
equipartition between the kinetic and Poynting energy fluxes at
large distances from the origin. In this case the Lorentz force
efficiently collimates the flow, which reaches cylindrical asymptotics.
In contrast, for $F<1$, even though the acceleration is more
efficient (so more of the Poynting flux is converted into
kinetic energy), the collimation is weaker and the flow only
reaches conical asymptotics.

\begin{figure*}[t]
\centerline{{\includegraphics[width=.97\textwidth]{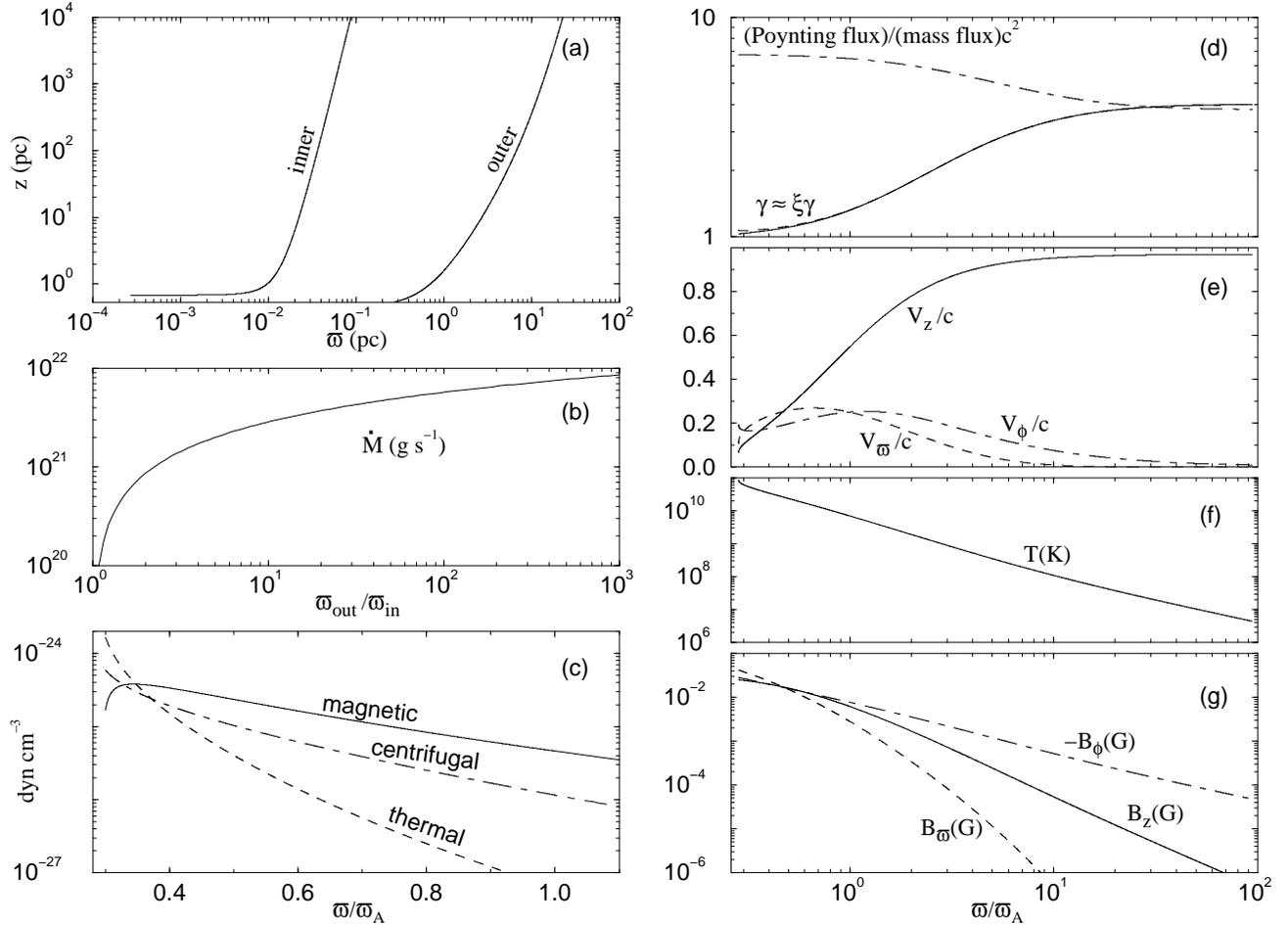}}}
\caption{
$r$ self-similar solution describing the jets in NGC 6251.
$(a)$ Poloidal field-line shape on a logarithmic scale.
$(b)$ Mass loss rate as a function of
${\varpi_{\rm out}}/{\varpi_{\rm in}}$,
the ratio of the outermost and innermost disk radii.
The remaining panels show the force densities in the poloidal
direction $(c)$ and various other quantities [$(d)-(g)$; see text
for details] as functions of
$\varpi/\varpi_{\rm A}$ 
(which, in turn, is a function of the polar angle $\theta$) along the outermost field line.
Here $\varpi_{\rm A}$ is the Alfv\'en lever arm, which equals
$\varpi_{\rm A, in} = 9.8\times 10^{-4}\ {\rm pc}$ and
$\varpi_{\rm A, out} = 10^3\, \varpi_{\rm A, in} = 0.98\ {\rm
pc}$ on the innermost and outermost
field lines, respectively. Along the innermost field line, the quantities 
plotted in panels $(d)$, $(e)$, and $(f)$ remain the same functions of 
$\varpi/\varpi_{\rm A}$, whereas the quantities shown in panels $(c)$ and $(g)$ are
$\varpi_{\rm A, out}/\varpi_{\rm A, in}$ times larger.
\label{fig1}}
\end{figure*}

The illustrative solutions derived in VK demonstrated that
centrifugal driving plays a limited role in the acceleration of
relativistic flows --- it is only important initially (for as long
as $V_\phi \gtrsim V_p$). As already noted in
\S~\ref{introduction}, thermal acceleration can dominate over a
more extended zone if the initial enthalpy is relativistic
($\xi_i \gg 1$). Magnetic acceleration, however, takes over
after $\xi$ drops to $\sim 1$ and can remain important well
beyond the classical fast-magnetosonic surface (where $\gamma
V_p \approx [(B^2-E^2)/(4 \pi \rho_0 \xi)]^{1/2}$). If the
Poynting flux is initially smaller than the enthalpy flux
(corresponding to $\mu \approx \xi_i$), the
acceleration is predominantly thermal (governed by $\xi\gamma
\approx \mu$) and, for a trans-Alfv\'enic solution, the magnetic
field only acts to guide the flow. The collimation in this case
is weak, and the streamlines are asymptotically conical \citep[see also][]{VPK03}.

\section{Magnetic Outflow Models for Jets in Radio Galaxies and Quasars}\label{magnetic}
\subsection{The Accelerating Jets in the Radio Galaxy NGC 6251}
\label{ngc6251}

Figure \ref{fig1} shows a solution of the steady,
axisymmetric, ideal-MHD equations that describes an outflow from
a disk around a supermassive black hole. As in VK, the solution
was constructed using the $r$ self-similarity formalism but was
restricted to extend over a finite radial range. The
dimensional parameter values were determined on the basis of the inferred magnitudes of the
NGC 6251 black-hole mass and mass outflow rate.\footnote{The
values of the model parameters and boundary conditions that define
the displayed solution are, in the notation of VK: $\Gamma=5/3$,
$F=1.0001$, $z_c=0.7\ {\rm pc}$, $x_{\rm A}^2=0.87$, $\sigma_{\rm M}=3$,
$\xi_{\rm A}=1.005$, $\mu=7.8$, $\theta_{\rm A}=50\degr$, and
$B_0 \varpi_0^{2-F}=1.1 \times 10^{16}\ {\rm cgs}$.}
Given that the slow-magnetosonic singular surface arises from
the interplay between gravitational and thermal forces and that
the relativistic $r$ self-similar model does not
incorporate gravity, the solution only covers the super-slow regime of the flow.
This regime, however, is the most pertinent one and --- even for the
high value of $T_i$ that we adopt (following \citealt{MLF02}) ---
contains the $V_p>0.13\, c$ velocity range measured in the
\citet{S00} observations.

Figure \ref{fig1}a depicts the field-line shape. The innermost field
line originates from the
vicinity of the black hole, at a (cylindrical) distance
$\varpi_{\rm in}$ of a few Schwarzschild radii from the center
(for a black hole of mass $M$, the Schwarzschild radius
is $\simeq 10^{-13}\ (M/ M_\sun)\ {\rm pc}$), whereas the outermost field line originates
at $\varpi_{\rm out} = 10^3\, \varpi_{\rm in}$.
For this choice of the $\varpi_{\rm out}/\varpi_{\rm in}$ ratio
and the inferred mass of the NGC 6251 black hole,
the mass-loss rate in the wind is $\simeq 9 \times 10^{21}\ {\rm
g\ s}^{-1}$ (see Fig. \ref{fig1}b), close to the best-fit estimate of \citet{MLF02}.

Figure \ref{fig1}c shows the various force densities in the
poloidal direction along the outermost field line as functions of
$\varpi/\varpi_{\rm A}$  (where $\varpi_{\rm A}\equiv
(L/\mu\Omega)^{1/2}$ is the Alfv\'en lever arm;
note that, in the $r$ self-similar model, $\varpi/\varpi_{\rm A}$ is solely
a function of the polar angle $\theta$). Although the
thermal pressure gradient is the dominant force density very close to the origin, 
the magnetic force rapidly takes over. (The centrifugal force,
which could in principle also contribute near the origin, is
much smaller than the pressure gradient force in this case.)
Figure \ref{fig1}d shows that the Lorentz factor ({\em
solid} line) increases monotonically with distance. Also shown
is the product $\xi \gamma$ ({\em dashed} line), which
demonstrates (in accord with Fig. \ref{fig1}c) that $\xi$
differs from 1 (signaling that thermal effects contribute to the
acceleration) only in the immediate vicinity of the origin
($\varpi \lesssim 0.35\, \varpi_{\rm A}$). It
is seen that, even for very high initial temperatures (Fig. \ref{fig1}f),
the thermal contribution to the acceleration is negligible,
confirming the conclusion of \S~\ref{thermalaccel}. The upper
curve in Figure \ref{fig1}d indicates that, for  $\varpi > 0.35\, \varpi_{\rm A}$,
the Lorentz factor increases due to a decreasing
Poynting-to-mass flux ratio: in this range the Poynting flux is
converted into matter kinetic-energy flux.
Asymptotically the Lorentz factor is $\simeq \mu /2$, or,
equivalently, the flow reaches a rough equipartition between 
Poynting and kinetic energy fluxes.
Although the final value of $\gamma$ in the jets in NGC 6251 is not
known, our particular choice of value for $\mu$ implies that $\gamma_\infty \approx 4$.
This estimate is consistent with the values typically inferred
for radio-galaxy jets \citep[e.g.,][]{G01}.

Figure \ref{fig1}e shows the various components of the flow velocity.
Although it is not crucial for a qualitative analysis, we note that the
solution presented here actually reproduces the velocities
inferred by \citet{S00}. Asymptotically $V_\varpi \ll V_z$,  as
expected in solutions that exhibit cylindrical collimation (see \S~\ref{model}).

The components of the magnetic field along the outermost field line 
are shown in Figure \ref{fig1}g.
The poloidal field initially exceeds the azimuthal
component (with the two becoming comparable at $\varpi = \varpi_{\rm A}$),
but at larger distances the azimuthal component dominates. 
The fields scale roughly as $B_\phi \propto \varpi^{-1}$ and $B_z \propto \varpi^{-2}$.

\subsection{The Accelerating Jet in the Quasar 3C 345}\label{3c345}

\begin{figure*}[t]
\centerline{
{\includegraphics[width=.97\textwidth]{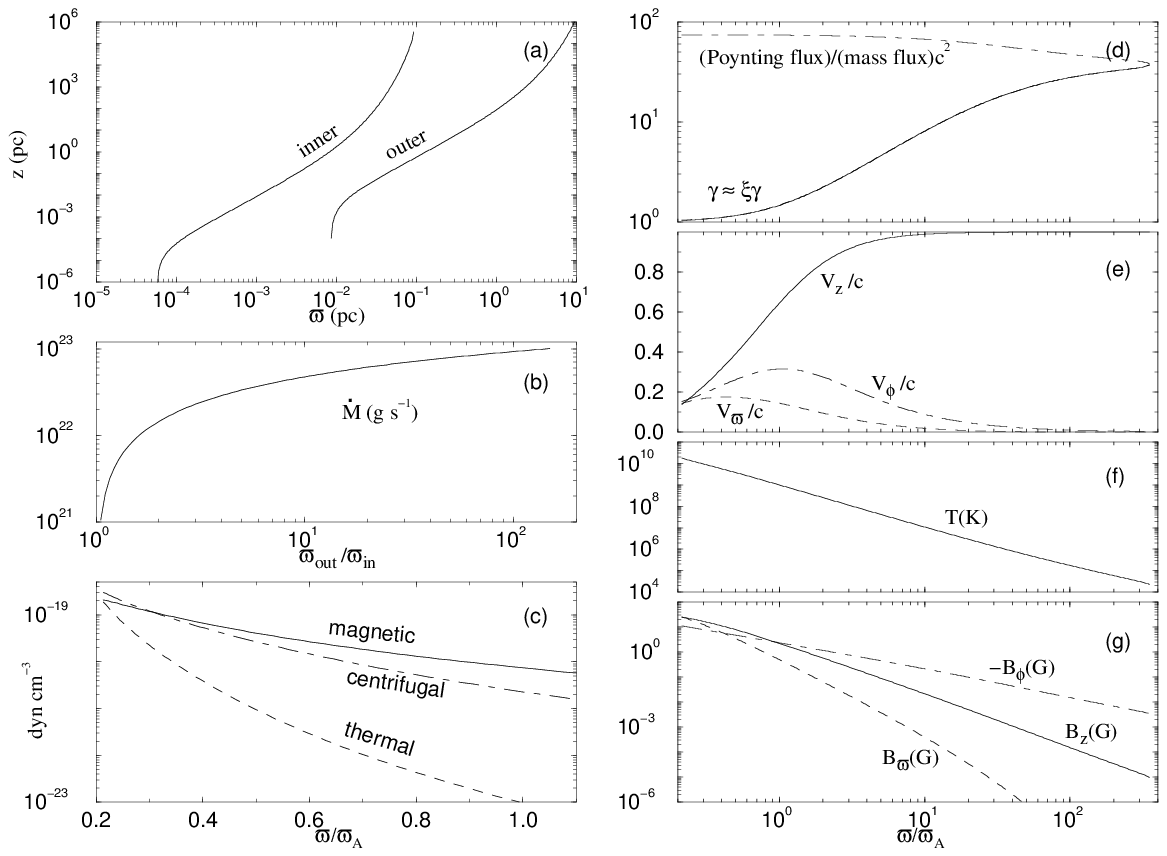}}}
\caption{
Same as Fig. \ref{fig1}, but for the 
application to the superluminal jet in 3C 345.
Here $\varpi_{\rm A, in} = 2.7\times 10^{-4}\ {\rm pc}$ and
$\varpi_{\rm A, out} = 150\, \varpi_{\rm A, in} = 4.1 \times 10^{-2}\ {\rm pc}$.
\label{fig2}}
\end{figure*}

We propose that the parsec-scale acceleration inferred by
\citet{U97} in component C7 of the 3C 345 jet is most plausibly
interpreted in terms of magnetic driving, and we present in Figure \ref{fig2} 
an $r$ self-similar MHD solution describing a
proton-electron jet that supports this claim.\footnote{
The parameters/boundary conditions that determine this solution
are, in the notation of VK: $\Gamma=5/3$,
$F=0.99$, $z_c=0$, $x_{\rm A}^2=0.987$, $\sigma_{\rm M}=8.66$, 
$\xi_{\rm A}=1.0007$, $\mu=75$,
$\theta_{\rm A}=25\degr$, and $B_0 \varpi_0^{2-F}=2.1 \times
10^{17}\ {\rm cgs}$.} In particular, it is seen from Figure
\ref{fig2}d that the Lorentz factor
increases from  $\gamma \approx 5$ at
$\varpi/\varpi_{\rm A} \approx 5.55$
to $\gamma \approx 10$ at $\varpi/\varpi_{\rm A} \approx 13.31$.
Using the value of the Alfv\'en lever arm on the
outer field line ($\varpi_{\rm A, out}=4.1 \times 10^{-2}\ {\rm pc}$),
we find that the cylindrical distance changes from
$\varpi\approx 0.23\ {\rm pc}$ to
$\varpi\approx 0.546\ {\rm pc}$; on the basis of Figure
\ref{fig2}a, these cylindrical radii correspond to linear
distances from the origin of $\sim 3\ {\rm pc}$ and $\sim 20\
{\rm pc}$, respectively, in close correspondence with the observed values.

The quasar jet solution shown in Figure \ref{fig2} is
characterized by a significantly higher value of the total
energy-to-mass flux ratio $\mu c^2$ than the radio-galaxy
solution depicted in Figure \ref{fig1}. The terminal Lorentz
factor, which again corresponds to a rough equipartition between
the asymptotic Poynting and kinetic energy fluxes  $(\gamma_\infty \approx
\mu/2)$ is correspondingly higher.
For the adopted fiducial parameters, the C7 component is
predicted to continue accelerating up to $\gamma_\infty \approx
35$. Interestingly, Lorentz factors of this order have been
inferred in the more distant components (in particular, C3 and
C5) of the 3C 345 jet \citep{LZ99}.

In comparison with the radio-galaxy solution of
\S~\ref{ngc6251}, the quasar solution presented in this subsection
corresponds to a more massive outflow (with a
mass-loss rate $\sim 10^{23}\ {\rm g\ s}^{-1}$ for $\varpi_{\rm
out}/\varpi_{\rm in}=150$; see Fig. \ref{fig2}b) and to a
stronger magnetic field (cf. Figs. \ref{fig1}g and \ref{fig2}g).
We note, however, that the density and magnetic field
strength are not uniquely determined from the kinematic data:
exactly the same flow speeds and field-line shape
are obtained if the density, particle pressure, and squared
amplitudes of the magnetic field components are rescaled by the
same factor.

\section{Summary and Discussion}
\label{conclusion}

We have argued that acceleration of AGN jets to relativistic
velocities on scales that are much larger than the
gravitational radius of the central black hole is most plausibly
explained in terms of magnetic driving.
This mechanism involves acceleration by the gradient of the
azimuthal magnetic-field pressure and is distinct from centrifugal
acceleration, which is often considered to be the dominant
driving mechanism of nonrelativistic jets. Centrifugal
driving takes place in the sub-Alfv\'enic flow regime and
accelerates the gas to a poloidal speed that is of the order of
the initial Keplerian speed in the underlying disk. In
comparison, magnetic pressure-gradient acceleration occurs over
a much more extended region (up to the modified
fast-magnetosonic surface) and can produce a much higher
(relativistic) terminal speed depending on the initial Poynting-to-mass 
flux ratio $\mu c^2$. In the trans-Alfv\'enic relativistic-MHD solutions
presented in this paper, the terminal Lorentz factor is
$\gamma_\infty \approx \mu /2$, corresponding to
a rough equipartition between the asymptotic Poynting and
kinetic-energy fluxes.\footnote{Note that the Lorentz factor on
the classical fast-magnetosonic surface is only $\sim
\mu^{1/3}$ \citep[e.g.,][]{C86}, so most of the acceleration in
these solutions occurs in
the super-fast regime.} These solutions are also characterized by strong magnetic
collimation (with the streamlines tending asymptotically to
cylinders) and are thus consistent with the narrow opening angles
inferred in AGN jets.

Thermal effects could in principle contribute to the
acceleration even in jets where a large-scale magnetic field
guides the flow. As discussed in VK, there are in general two thermal force densities:
the pressure gradient $-\nabla P$
and the ``temperature'' force\break
$-\gamma^2 \rho_0 \left({\boldsymbol{V}} \cdot \nabla \xi
\right) {\boldsymbol{V}} = -(\gamma^2 {\boldsymbol{V}} / c^2)
{\boldsymbol{V}}\cdot \nabla P$. These forces accelerate the
flow to $\gamma \approx \xi_i$. In cases where $\xi_i \gtrsim
1$, the thermal acceleration takes place
in the nonrelativistic regime and is terminated by the time the
speed increases to $\sqrt{3} C_{{\rm s}, i}$ --- i.e., just beyond the sonic surface
that typically lies very close to the origin. As pointed out in \S~\ref{thermalaccel},
this situation applies to proton-electron outflows; therefore,
to the extent that AGN jets have a dynamically dominant
proton component (as is often inferred to be the case), their
acceleration to relativistic speeds will not be significantly
influenced by thermal effects.\footnote{
In the case of an electron-positron outflow,
or under optically thick conditions when radiation pressure
contributes strongly to the specific enthalpy, one can have $\xi_i \gg 1$. 
In this case the thermal driving (dominated by the
``temperature'' force) acts well beyond the sonic surface and
accelerates the flow to a highly relativistic speed (see \S~\ref{introduction}). If, in
addition to $\xi_i\gg 1$, $\mu\gg \xi_i$ also holds, then
magnetic driving takes over at the end of the thermal
acceleration zone and eventually increases the Lorentz factor to
$\sim \mu$ (see \S~\ref{model}): this is the behavior
obtained in the GRB jet models referenced in footnote \ref{grbref}.}

Although we only considered two specific applications ---
sub--parsec-scale acceleration involving moderately relativistic speeds
in a radio-galaxy jet (NGC 6251; \S~\ref{ngc6251}) and parsec-scale
acceleration involving highly relativistic speeds in a superluminal radio
quasar (3C 345; \S~\ref{3c345}) --- there are already several other
reported cases of relativistic AGN jets that show evidence
for a parsec-scale acceleration. In some cases this has been
deduced from an increase in the apparent speed of a particular
superluminal component \citep[e.g.][]{H96}. In other cases,
where there are observations of several superluminal components,
it was found that the
innermost one typically exhibits the smallest proper motion,
with more distant components indicating an acceleration on parsec
scales \citep[e.g.,][]{H01a}. In the case of the quasar 3C 279 jet,
\citet{P03} inferred an acceleration from $\gamma = 8$ at $r< 5.8\ {\rm
pc}$ to $\gamma = 13$ at $r \approx 17.4\ {\rm pc}$ using a
similar approach to the one that had been employed by
\citet{U97} in 3C 345. It is also worth noting in this connection that a
variety of observations indirectly support the magnetic
acceleration picture for AGN jets. For example, the parsec-scale helical
field morphology implied by our model is consistent with
VLBI polarization maps of BL Lac objects (e.g., \citealp{G00})
and with circular-polarization measurements of blazars (e.g., \citealp{H01b}).
However, we defer a more detailed discussion of the additional
observational implications of this model to future publications
in this series.

In conclusion, we reemphasize that our modeling framework is
quite general and is potentially applicable to relativistic
jets in a variety of astrophysical settings. In our previous
application to GRBs, the model could account for the inferred
values of $\gamma_\infty$ and of the upper limit ($\sim 10^{14}\
{\rm cm}$) on the size of the acceleration region, but it could
not be further constrained because the acceleration region in
GRB sources is not resolved. In contrast, the motion of
the radio components in certain microquasar jets has been
monitored on scales $\sim 10^{16}\ {\rm cm}$ \citep[e.g.,][]{F99}. 
It would thus be interesting to search for evidence of extended
acceleration in these sources, in analogy with the situation in AGN jets.

\acknowledgements 
This work was supported in part by NASA grants NAG5-9063 and NAG5-12635.

\end{document}